\documentstyle[psfig,11pt,a4]{article}

\newcommand{\be}{\begin{equation}}
\newcommand{\ee}{\end{equation}}
\newcommand{\bea}{\begin{eqnarray}}
\newcommand{\eea}{\end{eqnarray}}

\begin{document}

\begin{titlepage}

\begin{flushright}
{\tt
    hep-th/0309160}
 \end{flushright}

\bigskip

\begin{center}

{\bf{\LARGE Generalized Virasoro anomaly \\
and stress tensor for dilaton coupled theories}}

\bigskip
\bigskip\bigskip
 A.  Fabbri$^a$, \footnote{e-mail: \sc
fabbria@bo.infn.it}
S. Farese$^b$ \footnote{\sc farese@ific.uv.es} and
 J. Navarro-Salas$^b$ \footnote{\sc jnavarro@ific.uv.es}

\end{center}

\bigskip%

\footnotesize
\noindent
 a) Dipartimento di Fisica dell'Universit\`a di Bologna and INFN
 sezione di Bologna, Via Irnerio 46,  40126 Bologna, Italy
 \newline
 b) Departamento de F\'{\i}sica Te\'orica and
    IFIC, Centro Mixto Universidad de Valencia-CSIC.
    Facultad de F\'{\i}sica, Universidad de Valencia,
        Burjassot-46100, Valencia, Spain.

\bigskip

\bigskip

\begin{center}
{\bf Abstract}
\end{center}

We derive the anomalous transformation law of the quantum stress
tensor for a 2D massless scalar field coupled to an external
dilaton. This provides a generalization of the Virasoro anomaly
which turns out to be consistent with the trace anomaly. We apply
it together with the equivalence principle to compute the
expectation values of the covariant quantum stress tensor on a
curved background. Finally we briefly illustrate how to evaluate
vacuum polarization and Hawking radiation effects from these
results.



\end{titlepage}

\newpage

Two-dimensional conformal invariance is a key ingredient to
understand critical behaviour of certain planar statistical
mechanical systems \cite{bpz}. It also plays a pivotal role in the
formulation of superstring theory \cite{polchinski} and in the
quantum mechanics of black holes. The Bekenstein-Hawking area law
is derived in many different ways by applying Cardy's formula for
conformal field theories living in the black hole horizon (see for
instance \cite{car} and references therein). The universal thermal
character of black hole radiation is also related to the fact that
matter fields exhibit two-dimensional conformal invariance in the
vicinity of the
horizon.\\

Many of the basic properties of 2d conformal field theories can be
obtained by studying a simple model, namely a massless scalar
field
 \be \label{mif}
S=-{1\over2}\int d^2x \left(\nabla f\right)^2\>. \ee Standard
canonical quantization and Wick theorem lead to the well-known
operator product expansion of the quantum (normal ordered) stress
tensor \bea T_{\pm\pm}(x^{\pm}) T_{\pm\pm}( x'^{\pm}) &=& \frac{1}{8\pi ^2
(x^{\pm}- x'^{\pm})^4}- \frac{1}{\pi (x^{\pm}- x'^{\pm})^2} T_{\pm\pm}( x'^{\pm})
\nonumber
\\ &-& \frac{1}{2\pi (x^{\pm}- x'^{\pm})}\partial_{\pm} T_{\pm\pm}( x'^{\pm}) +\
... \ , \eea where $x^{\pm}=x^0 \pm x^1$ are null Minkowskian
coordinates. The above expansion leads to the Lie algebra \bea
\left[  T_{\pm\pm}(x^{\pm}),  T_{\pm\pm}(x'^{\pm }) \right] &=&
\frac{1}{2\pi} \partial_{x^{\pm}}\delta(x^{\pm}-x'^{\pm })
T_{\pm\pm}(x'^{\pm }) -\frac{1}{96\pi^2}
\partial_{x^{\pm}}^3 \delta (x^{\pm} - x'^{\pm }) \nonumber \\ &-&
\left( x^{\pm}\leftrightarrow x'^{\pm} \right) \ .\eea Since
$T_{\pm\pm}(x^{\pm})$, up to normalization, are the generators of
infinitesimal conformal transformations $x^{\pm} \to x^{\pm}+ \epsilon
^{\pm}(x^{\pm})$, this implies the following infinitesimal transformation
law for the stress tensor  \be \label{actio}\delta_{\epsilon ^{\pm}}
T_{\pm\pm} =\epsilon ^{\pm}
\partial_{\pm} T_{\pm\pm} +2\partial_{\pm}\epsilon ^{\pm}
T_{\pm\pm}-\frac{1}{24\pi}\partial_{\pm}^3\epsilon ^{\pm}\ .\ee
Exponentiating the action (\ref{actio}) one gets, under the
conformal transformation $x^{\pm} \to y^{\pm}(x^{\pm})$, the following
anomalous transformation law \be \label{trl}  T_{\pm\pm}(y^{\pm}) =
\left( \frac{dx^{\pm}}{dy^{\pm}}\right)^2  T_{\pm\pm}(x^{\pm}) -
\frac{1}{24\pi}\{ x^{\pm}, y^{\pm} \} \ ,\ee where $\{ x^{\pm}, y^{\pm}\}=
\frac{\partial^3 x^{\pm}}{\partial y^{\pm 3}}/ \frac{\partial x^{\pm
}}{\partial y^{\pm}} -\frac{3}{2} \left(\frac{\partial^2 x^{\pm}}{\partial
y^{\pm 2}}/\frac{\partial x^{\pm }}{\partial y^{\pm}}\right)^2$ is the
Schwarzian derivative. All these expressions can be regarded as
different realizations of the so-called Virasoro anomaly. For a
generic conformal field theory the above results are valid
provided we multiply the $c$-number terms of the above equations
by the central charge $c$ characterizing the
particular theory \cite{bpz}. \\

The first aim of this work is to study the modification of the
transformation law (\ref{trl}), when a dilaton field $\phi$ is
present and (\ref{mif}) is replaced by \be \label{mifcd}
S=-{1\over2}\int d^2x  e^{-2\phi}\left(\nabla f\right)^2\>. \ee A
nice justification of the form of the dilaton coupling comes from
General Relativity. If a scalar field $f$ is minimally coupled to
a 4D spherically symmetric metric \be
ds^{2}_{(4)}=ds^{2}_{(2)}+e^{-2\phi}d\Omega^2\ , \ee and we
perform dimensional reduction from $-\frac{1}{8\pi }\int d^4
x\sqrt{-g}(\nabla f)^2, $
we obtain the above action (\ref{mifcd}) in case of flat 2d space.
We shall also study the quantum stress tensor of the theory (\ref{mifcd}) in a generic two-dimensional curved background.\\

Let us now consider a simple case, namely the one associated to
the four-dimensional Minkowski space. In this situation it is
$ds^{2}_{(2)}=-dx^+dx^-$, where $x^{\pm}=t\pm r$, and
$e^{-2\phi}=r^2$. The mode expansion of the field $f$ living in
the $t-r$ plane (with the condition $f(r=0)=0$) is \be
f=\int_{0}^{\infty} \frac{dw}{\sqrt{4\pi w}}\left[ a_w (e^{-iwx^+}
- e^{-iwx^-}) + a_w^\dagger (e^{iwx^+} - e^{iwx^-}) \right]
e^{\phi}\ .\ee The null components of the stress tensor are given
by \be T_{\pm\pm}(x^+,x^-)=e^{-2\phi}(\partial_{\pm} f)^2, \ee and the
corresponding normal ordered operators can be defined, as usual,
by point-splitting (from now on we shall use an explicit notation
for the normal ordered stress tensor) \be :T_{\pm\pm}(x^+,x^-): =
\lim_{x^{\pm}\to x'^{\pm }} e^{-(\phi(x) +\phi(x'))}
\frac{\partial}{\partial x^{\pm}}\frac{\partial}{\partial x'^{\pm }}
(f(x)f(x')- \left< f(x)f(x')\right>), \ee where the two-point
function is \be \left< f(x)f(x') \right> = -\frac{1}{4\pi}
e^{\phi(x)+\phi(x')}\ln\frac{ (x^{+ }-x'^{+})(x^- - x'^{-})}{(x^{+
}-x'^{-})(x^- - x'^{+})}\ .\ee Under a conformal transformation
$x^{\pm}\to y^{\pm}(x^{\pm})$ normal ordering breaks covariance and the
transformed stress tensor picks up the following anomalous
non-tensorial contributions \bea \label{trlm}
 :T_{\pm\pm}(y^+,y^-): &=& \left( \frac{dx^{\pm}}{dy^{\pm}}\right)^2
 :T_{\pm\pm}(x^+,x^-): -\frac{1}{24\pi}\{ x^{\pm}, y^{\pm} \}\nonumber \\
&-& \frac{1}{4\pi} \left[ \frac{d^2 x^{\pm}}{dy^{\pm
2}} \left(\frac{dx^{\pm}}{dy^{\pm}} \right)^{-1}
\frac{\partial\phi}{\partial y^{\pm}} + \ln \left( \frac{dx^+}{dy^+}
\frac{dx^-}{dy^-} \right)
\left( \frac{\partial\phi}{\partial y^{\pm}}\right)^2 \right] \ . \eea

This expression generalizes the Virasoro-type transformation law
(\ref{trl}) by adding terms depending on the derivatives of
$\phi$. At this point we would like to remark that the above
expression has been obtained for a particular form of $\phi$ in
terms of the null coordinates $x^{\pm}$, namely \be \label{fldi}
\phi=-\ln \frac{x^+ - x^-}{2} .\ee However we want to stress that
the result has general validity, irrespective of the particular
form of the external dilaton field. We shall prove
this in two different ways: \\
i) the short-distance behaviour for the Hadamard function  does not depend on the
specific model; \\
ii) we shall show that eq. (\ref{trlm}) is the only local expression which is
consistent with the trace anomaly derived in the context of
gravitational physics \cite{mukanov-wipf-zelnikov}.
\\

We point out that the conformal symmetry can be recovered in regions where
$\partial_{\pm}\phi \to 0$. This happens typically when $r$ approaches infinity
and also, in the context of curved spacetime, at the black hole horizons.
\\

The equation of motion for the field $f$, derived from the action
(\ref{mifcd}), is \be\label{eom} \partial_+(e^{-2\phi}\partial_-
f)+\partial_-(e^{-2\phi}\partial_+ f)=0. \ee In general this can
be solved only for particular forms of $\phi$, for instance in the
situation where it is given by (\ref{fldi}) or \be \phi =
-\frac{1}{2}\ln\frac{(x^+ - x^-)}{2}\ .\ee In the latter case the
equation of motion for $f$ (\ref{eom}) coincides with the equation of
a minimal scalar
field in a three-dimensional spacetime, described by the action
\be S=-\frac{1}{4\pi}\int d^3 x \sqrt{-g}(\nabla f)^2\ , \ee
under the assumption of
axi-symmetry for the field $f$ and the metric
$ds^2_{(3)}=ds^2_{(2)}+r^2 d\varphi^2 $,
  where the radial function is given by $r=e^{-2\phi}$.
This equation turns out to be equivalent to one equation of the
Einstein-Rosen subsector of pure General Relativity. The system
is exactly solvable both classically and quantum-mechanically
(details can be found in \cite{Kuchar},
\cite{Ashtekar-Angullo-Cruz}, \cite{Niedermaier}, \cite{Barbero})
and, therefore, it can provide a nontrivial test of the formula
(\ref{trlm}). The field $f$ can be expanded in modes as follows
\be f=\int_{0}^{\infty} \frac{dw}{\sqrt{2}} J_0(r w)\left[ a_w
e^{-iwt} +a^\dagger_w e^{iwt}\right] \ee where $J_0$ is the zero
order Bessel function. At the quantum level the coefficients $a_w$
and $a_w^\dagger$ are converted into annihilation and creation
operators obeying the commutation relation
$[a_w,a_{w'}^\dagger]=\delta(w-w')$. To work out the quantum
behaviour of the stress tensor we need to evaluate the Hadamard
function $G^{(1)}(x,x')\equiv \frac{1}{2}\left< 0|\{ f(x), f(x')\}
|0\right>$. This turns out to be equal to \cite{Barbero2},
\cite{Niedermaier}\\
i) for $0<|t'-t|<|r'-r|$

$$G^{(1)}(x,x')=\frac{1}{\pi\!\sqrt{[(r'\!+\!
\!r)^2-(t'\!-\!\!t)^2]}}
K\!\!\left(\!\!\sqrt{\!\frac{4rr'}{(r'\!+\!
\!r)^2-(t'\!-\!\!t)^2}}\right);$$ \\
ii) for $|r'-r|<|t'-t|<r'+r $
$$
G^{(1)}(x,x')=\frac{1}{2\pi}\frac{1}{\sqrt{ r
r'}}\,K\left(\sqrt{\frac{(r'\!+\!r)^2-(t'\!-\!t)^2}
{4rr'}}\right).
$$
iii) for $r+r'<|t'-t|$ it is $G^{(1)}(x,x')=0$,\\
where $K(k)=\int_0^{\pi/2} d\theta /\sqrt{1-k^2\sin^2(\theta)}$ is
the complete elliptic integral. Using the expansion \cite{GR}
\be K(k')= \ln\frac{4}{k'} + (\frac{1}{2})^2 \left(
\ln\frac{4}{k'} - 1\right) k'^{2} +O(k'^4 \ln\frac{4}{k'}), \ee where
$k'=\sqrt{1-k^2}$, we obtain \bea\label{haex}
G^{(1)}(x,x')&=&-\frac{e^{\phi(x)+\phi(x')}}{4\pi} [ \ln(x^{
+}-x'^+)(x^{-}-x'^-) + const. \nonumber \\ &+&
 O\left(  (x^{+}-x'^+)(x^{-}-x'^-)\ln(x^{+}-x'^+)(x^{-}-x'^-)\right) ]\ .\eea
In the computation of the transformation law of the stress tensor,
via point-splitting, only the leading term in (\ref{haex})
produces a nontrivial contribution. Therefore it is easy to see
that the final result is (\ref{trlm}). Moreover, the above
expression agrees with the De Witt-Schwinger expansion of
$G^{(1)}(x,x')$, restricted to flat space-time, given in
\cite{bunch-christensen-fulling},
 \cite{balbinot-fabbri-nicolini-frolov}
\be G^{(1)}(x,x')=\frac{e^{\phi(x)+\phi(x')}}{2\pi}\left[ -(\gamma
+\frac{1}{2}\ln\frac{m^2\sigma}{2}) +O(\sigma\ln\sigma) \right]\ ,\ee
where $\gamma$ is the Euler constant,  $m^2$ is an infrared cutoff
and  $\sigma$ is one half the square of the distance between the
points $x$ and $x'$.\\

Due to presence of $\phi$ the classical conservation laws $\partial_{\mp}
T_{\pm\pm}=0$ get modified to (see \cite{balbinot-fabbri}, \cite{kummer-vassilevich}
for a
higher-dimensional interpretation) \be \label{qucl}
\partial_{\mp} T_{\pm\pm} + \partial_{\pm}\phi \frac{\delta S}{\delta \phi}=0,\ee
where \be \frac{\delta S}{\delta\phi}= -2e^{-2\phi}\partial_+
f\partial_- f\ .\ee Let us analyze the quantum analogous of these
equations. The transformation law for $\left<  :T_{\pm\pm}:
\right>$ is given by eq. (\ref{trlm}) and the corresponding one
for $\left< \frac{\delta S}{\delta \phi} \right>$ should be, on
general grounds, of the form \be \label{prmo} \left< \frac{\delta
S}{\delta \phi} (y^{\pm}) \right> = \frac{dx^+}{dy^+}\frac{dx^-}{dy^-}
\left<  \frac{\delta S}{\delta \phi} (x^{\pm}) \right> +  \Delta (\phi
;x^{\pm},y^{\pm}).\ee Let us suppose that \be \label{clqu}
\partial_{\mp}\left< :T_{\pm\pm}:\right> +
\partial_{\pm}\phi  \left< \frac{\delta S}{\delta
\phi}\right>=0.\ee If we transform this relation according to
(\ref{trlm}) and (\ref{prmo}) we get, by consistency, \bea &-&
\frac{1}{4\pi}\frac{ \frac{\partial^2 x^{\pm}}{\partial y^{\pm 2}}}{
\frac{\partial x^{\pm}}{\partial y^{\pm}}} \frac{\partial}{\partial
y^+}\frac{\partial}{\partial y^-}\phi -\frac{1}{2\pi}\ln \left(
\frac{\partial x^+}{\partial y^+}  \frac{\partial x^-}{\partial
y^-}\right)  (\frac{ \partial\phi}{\partial
y^{\pm}})\frac{\partial}{\partial y^+} \frac{\partial}{\partial
y^-}\phi \nonumber \\ &-& \frac{1}{4\pi} \left( \frac{\partial
\phi}{\partial y^{\pm}}\right)^2 \frac{ \frac{\partial^2
x^{\mp}}{\partial y^{\mp 2}}}{ \frac{\partial x^{\mp}}{\partial
y^{\mp}}} + \frac{\partial \phi}{\partial y^{\pm}} \Delta (\phi
;x^{\pm},y^{\pm}) =0 .\eea These two equations are compatible with the
uniqueness of $\Delta(\phi ;x^{\pm},y^{\pm})$ only if \be \label{requ}
\Box\phi=(\nabla\phi)^2.\ee If $\phi$ does not obey (\ref{requ})
the quantum conservation law (\ref{clqu}) must be modified. We
find that the only possibility to maintain consistency with the
transformation law (\ref{trlm}) is by adding a nontrivial trace
$\left< T_{+-}\right>$ just of the form \be \label{fsan} \left<
T_{+-}\right> = -\frac{1}{4\pi}\left(
\partial_+\phi \partial_-\phi -
\partial_+\partial_-\phi \right).\ee
Then for $\Delta(\phi ;x^{\pm},y^{\pm})$ we obtain \bea
\label{anomalousdelta} \Delta &=& \frac{1}{2\pi}\ln \left(
\frac{dx^+}{dy^+}\frac{dx^-}{dy^-}\right) \frac{d^2
\phi}{dy^+dy^-} \nonumber \\ &+&\frac{1}{4\pi}\left[ \frac{d^2
x^-}{dy^{- 2}}\left( \frac{dx^-}{dy^-} \right)^{-1}
\frac{d\phi}{dy^+}  + \frac{d^2 x^+}{dy^{+ 2}}\left(
\frac{dx^+}{dy^+} \right)^{-1} \frac{d\phi}{dy^-} \right] \ .\eea
 Finally, the quantum
conservation law, invariant under conformal transformations, reads
\be \label{qcl}
\partial_{\mp}\left< :T_{\pm\pm}:\right> + \partial_{\pm} \left< T_{+-}\right>
+\partial_{\pm}\phi \left< \frac{\delta S}{\delta \phi}\right> =0. \ee
We have to point out that the anomalous trace derived in this
approach agrees with the one derived in curved space-time (first
derived in  \cite{mukanov-wipf-zelnikov}) . For the
dilaton-coupled theory the trace anomaly, obtained in a covariant
quantization scheme, is \be \label{tranom} \left< T \right> =
\frac{1}{24\pi}\left( R - 6(\nabla\phi)^2 + 6\Box\phi \right) .\ee
If we restrict to flat space-time we obtain (\ref{fsan}). We
mention that in \cite{bouha} the above trace anomaly was derived
with a different numerical coefficient for the $\Box\phi$ term.
This coefficient was then corrected, according to (\ref{tranom}) ,
in \cite{corr} (the same result was obtained in \cite{molti}) . So
our derivation can be seen, as a by-product, as an alternative and
simple way to get the dilaton contribution to the trace anomaly.
Moreover, the argument can be applied the other way around:
assuming (\ref{fsan}), (\ref{qcl}) and locality one gets the
$\phi$ dependent terms of (\ref{trlm}).\\

Now we shall apply the above results to gravitational physics. We
shall work out an expression for the expectation values of the
covariant stress tensor using the anomalous transformation law
(\ref{trlm}) and the help of the equivalence principle to deal
with curved space. In a generic point $X$ of the space-time one
can always introduce locally inertial coordinates
$\xi_{X}^{\alpha}$. Restricting our attention to the
$(t-r)$-sector we can then construct the corresponding null
coordinates $\xi_{X}^{\pm}$. Since normal ordering breaks general
covariance we need a different prescription to construct a quantum
stress tensor compatible with diffeomorphism invariance. One can
do it starting from the expectation value of the normal ordered
stress tensor $\langle \Psi | T_{\pm\pm}(\xi^{\pm}(X))|\Psi\rangle$ in
the locally inertial frame $\{ \xi^{\pm}_X \}$ with respect to some
generic state $|\Psi\rangle$. The corresponding expectation values
in the curved background, at the generic point $X$ in the
coordinates $\{ x^{\pm}\}$, can be naturally defined as \be
\label{qstdilaton1}
\langle\Psi|T_{\pm\pm}(x^{+}(X),x^-(X))|\Psi\rangle \equiv
\left(\frac{d{\xi}_{X}^{\pm}}{dx^{\pm}}(X)\right)^2
\langle\Psi|:T_{\pm\pm}({\xi}_{X}^{+}(X),
{\xi}_{X}^-(X):|\Psi\rangle \ ,\ee this way we get the desired
covariant property
 \be\label{definitionqst2}
\langle\Psi|T_{\pm\pm}(y^{+}(X),y^-(X))|\Psi\rangle =
\left(\frac{dx^{\pm}}{dy^{\pm}}(X)\right)^2\langle\Psi|T_{\pm\pm}(x^{+}(X),x^-(X))|\Psi\rangle
 \ee
where $\{y^{\pm}\}$ and $\{x^{\pm}\}$ are arbitrary coordinate systems
around the generic point $X$.

 Now the relation between $:
T_{\pm\pm}(x^+(X),x^-(X)):$ and $:
T_{\pm\pm}(\xi^+_X(X),\xi^-_X(X)):$ is given by (using
(\ref{trlm})):\bea \label{trlm2}
 & & :T_{\pm\pm}(x^+(X),x^-(X)): =  \left( \frac{d\xi_X^{\pm}}{dx^{\pm}}(X)\right)^2
 :T_{\pm\pm}(\xi_X^+(X),\xi_X^-(X)):
 \nonumber \\ &-& \frac{1}{24\pi}\{ \xi_X^{\pm}, x^{\pm} \}|_X
 - \frac{1}{4\pi} [ \frac{d^2 \xi_X^{\pm}}{dx^{\pm
2 }}(X) \left(\frac{d\xi_X^{\pm}}{dx^{\pm}}(X) \right)^{-1}
\frac{d\phi}{d x^{\pm}}(X) \nonumber \\ &+& \ln
\frac{d\xi_X^+}{dx^+}(X)\frac{d\xi_X^-}{dx^-}(X) \left(
\frac{d\phi}{d x^{\pm}}(X) \right)^2 ] \ . \eea

Inserting (\ref{trlm2}) into (\ref{qstdilaton1}) we finally obtain
\bea \label{qstdilaton2} & &
\langle\Psi|T_{\pm\pm}(x^{+}(X),x^-(X))|\Psi\rangle =
\langle\Psi|:T_{\pm\pm}(x^{+}(X),x^-(X)):|\Psi\rangle \nonumber \\
&+& \frac{1}{24\pi}\{ \xi_X^{\pm}, x^{\pm} \}|_X
 + \frac{1}{4\pi} [ \frac{d^2 \xi_X^{\pm}}{dx^{\pm
2 }}(X) \left(\frac{d\xi_X^{\pm}}{dx^{\pm}}(X) \right)^{-1}
\frac{d\phi}{d x^{\pm}}(X) \nonumber \\
&+& \ln \frac{d\xi_X^+}{dx^+}(X)\frac{d\xi_X^-}{dx^-}(X) \left(
\frac{d\phi}{d x^{\pm}}(X) \right)^2 ] \ . \eea

To go further we need the relations between $\{ \xi_X^{\pm} \}$ and
$\{ x^{\pm} \}$. Up to second order and Poincar\'e
 transformations they are unambiguous
  and can be chosen to be conformal \cite{weinberg}
 \be
 \xi^{\pm}_X = b^{\pm}_{\pm}\left[ (x^{\pm} - x^{\pm}(X)) +
 \frac{\Gamma^{\pm}_{\pm\pm}}{2}(x^{\pm} - x^{\pm}(X))^2
 + F_{\pm}(x^{\pm}-x^{\pm}(X))^3  + ... \right] \ .\ee
 In a conformal frame $ds^2=-e^{2\rho}dx^+dx^-$
 the constants $b^{\pm}_{\pm}$ satisfy the constraint $b^+_+b^-_-=e^{2\rho(X)}$
 and $\Gamma^{\pm}_{\pm\pm}=2\partial_{\pm}\rho$.
 Note that the Schwarzian derivative requires the third order as well,
 which is not determined by the requirement that $\xi^{\pm}_X$ are locally inertial.
 We naturally fix it by imposing that, for a flat
 metric, $\xi^{\pm}(X)$ are the global null minkowskian coordinates. This leads to
 \be
 F_{\pm}=  \frac{1}{3}\partial_{\pm}^2\rho(X) + \frac{2}{3}\left(
 \partial_{\pm}\rho(X)\right)^2  \ .\ee

Using now the above expressions  a straightforward computation
  leads to the following form for the
  stress tensor, for an arbitrary point $X$,
\bea \label{qstdilaton3}
\langle\Psi|T_{\pm\pm}(x^+,x^-)|\Psi\rangle &=&
\langle\Psi|:T_{\pm\pm}(x^+,x^-):|\Psi\rangle
-\frac{1}{12\pi}(\partial_{\pm}\rho\partial_{\pm}\rho -
\partial_{\pm}^2\rho) \nonumber \\ &+& \frac{1}{2\pi} \left[  \partial_{\pm}\rho
\partial_{\pm}\phi + \rho (\partial_{\pm}\phi)^2 \right]\ .
\eea We remark that neglecting the terms containing the dilaton
these are the null components of the stress tensor derived from
the Polyakov effective action \cite{polyakov}.

 We want to compute now a covariant expression for $\langle \frac{\delta
S}{\delta \phi} \rangle $. To this end we shall impose the quantum
covariant conservation laws \be \nabla^{\mu}\langle T_{\mu\nu}
\rangle = \nabla_{\nu}\phi \frac{1}{\sqrt{-g}}\langle \frac{\delta
S}{\delta \phi}\rangle \ ,\ee which in the conformal frame are
translated into \be \label{qccl}
\partial_{\mp}\left< T_{\pm\pm}\right> + \partial_{\pm} \left< T_{+-}\right>
-2\partial_{\pm}\rho \langle T_{+-}\rangle +\partial_{\pm}\phi \left<
\frac{\delta S}{\delta \phi}\right> =0. \ee The $\langle T_{+-}
\rangle$ component is, as usual, fixed by the trace anomaly: \be
\label{tracedilaton+-} \langle T_{+-} \rangle =-\frac{1}{12\pi}
\left(
\partial_+\partial_-\rho + 3\partial_+\phi\partial_-\phi
-3\partial_+
\partial_-\phi \right) . \ee Combining
(\ref{tracedilaton+-}), (\ref{qstdilaton3}) and (\ref{qccl}) the
final result is \be \label{pressurecovariant} \langle \Psi|
\frac{\delta S}{\delta \phi}|\Psi  \rangle = \langle
\Psi|\frac{\delta S}{\delta \phi}|\Psi \rangle _{\rho=0}
-\frac{1}{2\pi} (
\partial_+\partial_-\rho +\partial_+\rho \partial_-\phi + \partial_-\rho \partial_+\phi
+2\rho \partial_+\partial_-\phi )\ . \ee The last three terms can
be obtained from the anomalous transformation law for $\langle
\frac{\delta S}{\delta \phi} \rangle _{\rho=0}$ (eqs.(\ref{prmo})
and (\ref{anomalousdelta})), while the term
$\partial_+\partial_-\rho$ comes directly from the imposition of
the conservation equations (\ref{qccl}). The state dependent
quantities in (\ref{qstdilaton3}) and (\ref{pressurecovariant})
are conserved, namely they satisfy \be \label{conslawflatspace}
\partial_{\mp}\langle\Psi|:T_{\pm\pm}(x^+,x^-):|\Psi\rangle +\partial_{\pm} \langle T_{+-}
\rangle |_{\rho=0} + \partial_{\pm}  \phi \langle \Psi|\frac{\delta
S}{\delta \phi}|\Psi \rangle _{\rho=0}=0\ . \ee This is a crucial
ingredient in order to fulfill equations (\ref{qccl}). To match
with the standard notation of 2D dilaton gravity \cite{Strominger}
we define the following functions $t_{\pm} (x^+,x^-)$ and
$t(x^+,x^-)$: \bea -\frac{1}{12\pi}t_{\pm}(x^+,x^-) &\equiv &
\langle\Psi|:T_{\pm\pm}(x^+,x^-):|\Psi\rangle \ , \nonumber \\
-\frac{1}{2\pi} t(x^+,x^-) &\equiv & \langle  \Psi|\frac{\delta
S}{\delta \phi}|\Psi \rangle _{\rho=0}\ \eea characterizing the
quantum state $|\Psi \rangle$. Notice that now, in contrast with
the minimally coupled case, the functions $t_{\pm}$ are no more chiral
(the same is true for the new function $t$) and satisfy a more
involved set of equations  reflecting the nontriviality of the
theory even in flat 2d space.

As an application of these equations we shall  perform a brief
analysis of the different choices of quantum states. To this end
let us consider the eternal Schwarzschild spacetime, described by
the 2d metric:
 \be
ds^2_{(2)}=-(1-2M/r)dudv\ , \ee  where $v=t+r^*$ and $u=t-r^*$,
$r^*=r+2M\ln (\frac{r}{2M}-1)$, and the dilaton field given by \be
e^{-2\phi}= r(u,v)^2\ . \ee We can naturally choose the state such
that ($\{ x^+=v, x^-=u\}$) \be \label{evlv}  t_{\pm} = 0 \ee and
consequently, because of (\ref{conslawflatspace}), \be
\label{functt}  -\frac {1}{2\pi}t \equiv \langle \Psi|\frac{\delta
S}{\delta \phi}|\Psi \rangle _{\rho=0}= -\frac{3
}{8\pi}\frac{M}{r^3} + \frac{1}{\pi}\frac{M^2}{r^4} \ . \ee This
corresponds to (or at least is a good approximation of) the
Boulware vacuum state $|B \rangle $ \cite{boulware}, describing
the vacuum polarization outside a static (not collapsed) star.
Applying expressions (\ref{qstdilaton3}) we get \be
\label{psistate} \left< B| T_{\pm\pm}|B \right> =
\frac{1}{24\pi}\left( -\frac{4M}{r^3}+
\frac{15}{2}\frac{M^2}{r^4}\right) + \frac{1}{16\pi r^2}
(1-\frac{2M}{r})^2 \ln (1-\frac{2M}{r}) \ . \ee The $+-$ component
is state independent and fixed by the trace anomaly (\ref{tranom})
\be \label{boulsta2} \left< T_{+-} \right>=
\frac{1}{12\pi}(1-\frac{2M}{r})\frac{M}{r^3}\ .\ee Finally, we
also have, from equations (\ref{pressurecovariant}) and
(\ref{functt}), \be \langle B|\frac{\delta S}{\delta \phi}|B
\rangle = -\frac{7}{8\pi} \frac{M}{r^3} +\frac{2}{\pi}
\frac{M^2}{R^4} + \frac{1}{8\pi
r^2}(1-\frac{4M}{r})(1-\frac{2M}{r})\ln (1-\frac{2M}{r})\ .\ee
Similar results, based on exact properties of the effective action
under Weyl transformations, were derived in
\cite{balbinot-fabbri2}. It is worthwhile to remark that in the
horizon limit and at infinity they are in agreement with the
results derived from canonical quantization
\cite{balbinot-fabbri-nicolini-frolov}.

A physically more interesting case is the one leading to black
hole evaporation. For it a natural choice is \be
-\frac{1}{12\pi}t_-(x^+,x^-)_{v \to -\infty} \sim \frac{1}{768\pi
M^2}\  \ee
 at the past horizon and
\be -\frac{1}{12\pi}t_+(x^+,x^-)_{u \to -\infty} \sim 0 \  \ee at
past null infinity. These conditions define the Unruh vacuum state
\cite{unruh}. In the absence of dilaton (minimally coupled
theory), the $t_{\pm}$ functions are chiral and then $t_{-}=-
\frac{1}{64 M^2}$ and $t_{+}=0$ everywhere. In terms of the Fock
space these conditions are related to the following density matrix
\be \rho_{U} = \prod_{w}\left( 1-e^{-2\pi w{\kappa}^{-1}} \right)
\sum_{{\stackrel{\rightarrow}{n}}}
e^{-2\pi{\stackrel{\rightarrow}{n}}w{\kappa}^{-1}}
|{\stackrel{\rightarrow}{n}_w}><{\stackrel{\rightarrow}{n}_w}| \ ,
\ee where  $|{\stackrel{\rightarrow}{n}_w}>$ is the state in the
Fock space with ${\stackrel{\rightarrow}{n}_w}$ outgoing particles
of frequency $w$. Without dilaton the corresponding modes are
plane waves and one can see immediately that this state reproduces
the above value for the function $t_{-}$ and, at future null
infinity, leads to the Hawking flux \be \label{hafu} \left<
U|T_{uu}|U \right>_{r \to +\infty} \sim \frac{1}{2\pi}
\int_{0}^{\infty} \frac{wdw}{e^{8\pi Mw}-1}= \frac{\pi}{6}T_H^2\ ,
\ee where $T_H = \frac{1}{8\pi M}$ is the Hawking temperature. In
the presence of the dilaton the modes are no longer planewaves
because they are affected by the potential barrier
\cite{balbinot-fabbri-nicolini-frolov}, \cite{De-Witt}. In this
case the result is \be \left< U|T_{uu}|U \right>_{r \to +\infty}
\sim \frac{1}{2\pi} \int_{0}^{\infty} \frac{wdw}{e^{8\pi
Mw}-1}|B(w)|^2= \xi \frac{\pi}{6}T_H^2 \ , \ee where $B(w)$ is the
transmission coefficient \cite{De-Witt} and $\xi$ the greybody
factor. The greybody factor $\xi$, related to $|B(w)|^2$ from the
above equation, produces a damping of the Hawking flux with
respect to that obtained without the dilaton coupling (for the
present theory it is $\xi\simeq 1.62/10$, see
\cite{balbinot-fabbri-nicolini-frolov}). For the massless
minimally coupled 2d scalar field there is no potential barrier,
hence $\xi=1$ and the Hawking flux is therefore given by
(\ref{hafu}). The evaluation of the expectation value of the
stress tensor at the future horizon also provides the expected
result. For the normal ordered operator we have \be \left<
U|:T_{vv}:|U \right>_{r \to 2M} \sim \frac{1}{2\pi}
\int_{0}^{\infty} \frac{wdw}{e^{8\pi Mw}-1}|A(w)|^2 \ , \ee where
$A(w)$ is the reflection coefficient. Now taking into account that
$|A(w)|^2+|B(w)|^2 =1$ and (\ref{qstdilaton3}) we get \be \left<
U|T_{vv}|U \right>_{r \to 2M} \sim -\frac{1}{2\pi}
\int_{0}^{\infty} \frac{wdw}{e^{8\pi Mw}-1}|B(w)|^2 \ . \ee We see
that this is the negative flux entering the black hole horizon
which compensates the Hawking radiation at infinity. A similar
study can also be performed for the Hartle-Hawking thermal
state.\\

 With the above analysis
concerning the choice of states we have checked again the physical
consistency of the proposed expression for the covariant quantum
stress tensor for dilaton coupled theories. However we have to
point out that the advantage of having the entire expression for
the quantum stress tensor is that it allows to properly consider
the one-loop semiclassical equations and to attack the interesting
and difficult problem of backreaction.

To end the paper, we would like to remark that the fact that
(\ref{trlm})  is the exact transformation law of the quantum
stress tensor for a generic dilaton field $\phi$ should not be a
surprise at all. One of the main features of 2d conformal field
theories is the existence of universal behaviours, irrespective of
the particular model considered. Therefore one could be tempted to
conjecture that (\ref{trlm}) is also valid for an arbitrary
conformal field theory coupled to a dilaton, up to numerical
coefficients in the $c$-numer terms.

 \section*{Acknowledgements}
This research has been partially supported by the research grants
BFM2002-04031-C02-01 and BFM2002-03681 from the Ministerio de
Ciencia y Tecnologia (Spain), EU FEDER funds and the INFN-CYCIT
Collaborative Program. S.F. acknowledges the Ministerio de
Educacion, Cultura y Deporte for a FPU fellowship. J. N-S would
like to acknowledge the Department of Physics of the University of
Bologna for hospitality. A.F. and J. N-S thank R. Balbinot and S.
Fagnocchi for useful discussions. Finally, we wish to thank the
authors of \cite{Barbero} for assistance concerning the 2+1 model
considered.

\end{document}